\documentstyle{article}

\author{L.S.F. Olavo\\
Departamento de Fisica, Universidade de Brasilia - UnB\\
70910-900, Brasilia - D.F. - Brazil}
\title{Quantum Mechanics as a Classical Theory \\
VI: The Classical Spin
}
\date{20, september, 1995
}

\begin{document}

\maketitle
\begin{abstract}
In these continuation papers (VI and VII) we are interested in approach the
problem of spin from a classical point of view. In this first paper we will
show that the spin is neither basically relativistic nor quantum but
reflects just a simmetry property related to the Lie algebra to which it is
associated. The classical approach will be paraleled with the usual quantum
one to stress their formal similarities and epistemological differences. The
important problem of Einstein-Bose condensation for fermions will also be
addressed.
\end{abstract}

\section{General Introduction}

The following two papers deals with the idea of classical spin. They are
intended to show that the concept of half-integral spin might be raised also
in the realm of a classical theory.

This first paper begins by showing that with some supposition about the
half-integer spin particle {\em structure} it is possible to recover all the
results derived using quantum theory. In this case, the language of
operators is substituted by a language of functions and the commutator is
replaced by the Poisson bracket. Functions formally identical with creation
and annihilation operators are defined and their interpretation---easier in
this formalism---is described. Another function related with the `number'
operator is also derived and interpreted. When passing from the passive to
the active view, the above mentioned functions become operators and we
recover again all the quantum mechanical formalism in Heisemberg's
notation---only the half-spin case is explicit treated.

The Exclusion Principle of Pauli is also addressed. It is shown that the
model assumed for the half-integral spin particle structure and the
imposition upon the eigenfunctions to make diagonal the operators $\widehat{S%
}_3$ and $\widehat{S}^2$ are sufficient to derive the exclusion behavior;
which now becomes a theorem.

As a secondary result of the picture proposed, it is shown that a
Bose-Einstein condensation for fermions might be easily interpreted by this
theory.

In the second paper, the phase-space expressions for the spin functions $S_3$
and $S^2$ are used to derive their equivalent quantum Schr\"odinger
equation. This equation is then solved and we find all the possible
half-integral spin eigenvalues and eigenvectors.

We then show quantitatively, why one expects fermionic Bose-Einstein
condensation to take place.

We end this series (VI and VII) with our final conclusions.

\section{Introduction}

Since its manifestation as a result of the Stern-Gerlach experiment---among
others---, the spin has been considered as a purely quantum mechanical
manifestation of Nature that has no resemblance to any classical behavior.
Many physicists defending an epistemological abyss between classical and
quantum worlds (or ontologies) are now accustomed to cite this effect, the
electronic spin, in support to their philosophical views\cite{varios}.

The aim of the present paper is to show that the electronic spin could also
be predicted from a purely classical point of view, without any reference to
quantum mechanics. This procedure is coherent with our previous developments
\cite{eu1-5,eu2-5,eu3-5,eu4-5,eu5-5}.

The main argument here is the fact that, when deriving the electronic spin,
one needs not to make reference to any of the quantum postulates. What one
needs to do is to use the Lie algebra induced by the product

\begin{equation}
\label{1}\left[ S_i,S_j\right] =i\hbar \epsilon _{ijk}S_k,
\end{equation}
where the $S_i$ are operators and $[,]$ represents the commutator of these
operators, together with some guess about the phenomenon itself revealed by
the Stern-Gerlach experiment.

As everybody knows, the same algebra is induced by the product\cite
{goldstein}

\begin{equation}
\label{2}\left\{ S_i,S_j\right\} =\epsilon _{ijk}S_k,
\end{equation}
where now the $S_i$ are functions and $\{,\}$ represents the Poisson
brackets.

In the next section we will see that with this product and some {\em picture}
or model of the electron internal structure it is possible to derive exactly
the same results already derived using the usual formalism of quantum
mechanics.

\section{The Classical Spin}

We begin this section by making a {\em picture} of the electron structure.
This procedure, of course, might be sustained only under a classical theory
since, as is amply known, the orthodox interpretation of quantum mechanics
does not allow us to make world pictures.

This picture of the electron structure will be done based on the expected
final results revealed by the Stern-Gerlach experiment, that is, space
``quantization''. The resemblances and differences between this approach and
the usual quantum mechanical one will be stressed whenever needed.

We want the electron to be a flat body possessing charge $e$ and mass $m$
that might have its geometry distorted in the plane perpendicular to its
symmetry axis and which is rotating in this plane. This structure for the
electron implies that we shall have
\begin{equation}
\label{cond1}z=p_z=0
\end{equation}
if we choose the $z$-axis to be the symmetry axis fixed on the electron
around which it is rigidly rotating (see fig. 1).

This geometry implies that the electron is capable of interacting with an
external magnetic field ${\bf B}$ with the interaction energy given \cite
{jackson} by

\begin{equation}
\label{3}H=-\frac e{2mc}{\bf B}\cdot {\bf L}
\end{equation}
where ${\bf L}$ is the angular momentum related to the electron
rotation---we forget for a moment the question about the Land\'e g-factor to
which we return soon.

The three quantities characterizing the electron are:

\begin{itemize}
\item  The angular momentum in the z-direction related with its rotation: $%
L_z$;
\begin{equation}
\label{4}L_z=xp_y-yp_x;
\end{equation}
with $L_x=L_y=0$ because of (\ref{cond1}).

\item  The quadrupole moments related to its possible distortions: $Q_{xy}$
and $Q_1$ where
\begin{equation}
\label{5}Q_{xy}=\sqrt{\frac \alpha \beta }xy+\sqrt{\frac \beta \alpha }p_xp_y%
\mbox{ and }Q_1=\frac 12\left[ \sqrt{\frac \alpha \beta }\left(
x^2-y^2\right) +\sqrt{\frac \beta \alpha }\left( p_x^2-p_y^2\right) \right]
\end{equation}
where all other quadrupole moments vanish identically because of the
condition $z=p_z=0$ and $\alpha $ and $\beta $ are structure dimensional
constants.
\end{itemize}

We now impose that the description of the electron be made in terms of the
three quantities $S_i,\ i=1,2,3$ related to the above defined moments as

\begin{equation}
\label{6}S_3=\frac 12L_z\mbox{ ; }S_2=\frac 12Q_{xy}\mbox{ ; }S_1=\frac
12Q_1.
\end{equation}

With this imposition it is easy to demonstrate that the three functions $S_i$
obey

\begin{equation}
\label{7}\left\{ S_i,S_j\right\} =\epsilon _{ijk}S_k,
\end{equation}
where we will use the classical Poisson Bracket throughout.

We also see that the functions
\begin{equation}
\label{8}S_0=\frac 12\left[ \sqrt{\frac \alpha \beta }\left( x^2+y^2\right)
+ \sqrt{\frac \beta \alpha }\left( p_x^2+p_y^2\right) \right]
\end{equation}
and

\begin{equation}
\label{9}S^2=S_1^2+S_2^2+S_3^2=\frac 1{16}.\left[ \sqrt{\frac \alpha \beta }%
\left( x^2+y^2\right) +\sqrt{\frac \beta \alpha }\left( p_x^2+p_y^2\right)
\right] ^2=\frac{S_0^2}4
\end{equation}
commute with all the $S_i,\ i=1,2,3$ (expression (\ref{9}) was expected
since, as we will see, the functions $S_i$ are the coordinate-momentum
representation of the SU(2) which is of rank one).

In terms of these functions, the interaction hamiltonian becomes
\begin{equation}
\label{10}H=-(2\omega _0)S_3=-(g\omega _0)S_3=-\omega _1S_3.
\end{equation}
where we have put the magnetic field in the direction of the $z$-axis to
write
\begin{equation}
\label{10.a}\omega _0=\frac{eB_3}{2mc}
\end{equation}
and call $g=2$ the Land\'e factor.

The equations of motion for these functions are related with the dynamics of
the problem. They might be obtained as usual by the use of the Poisson
Bracket and are given by

\begin{equation}
\label{11} \frac{dS_1}{dt}=\left\{S_1,H\right\}=\omega_1 S_2(t),
\end{equation}
and
\begin{equation}
\label{12} \frac{dS_2}{dt}=-\omega_1 S_1(t)\mbox{ ; }\frac{dS_3}{dt}=0.
\end{equation}

If we integrate these equations we find
$$
S_1(t)=S_1(0)\cos(\omega_1 t)+S_2(0)\sin(\omega_1 t);
$$

\begin{equation}
\label{13}S_2(t)=-S_1(0)\sin (\omega _1t)+S_2(0)\cos (\omega _1t)\mbox{ ; }%
S_3(t)=S_3(0).
\end{equation}
Equations (\ref{13}) represent a precession taking place in the space
defined by the functions $S_i$. This precession shall not be confused with
one in three-dimensional space, since in this space we postulate from the
very beginning that the rotation is taking place rigidly along the symmetry
axis. Instead, we might interpret this motion as a vibration of the electron
structure over the $xyp_xp_y$ phase space hyperplane where it can be
distorted, since the functions $S_1$ and $S_2$ are related with the
quadrupole moments that take into account such a distortion (fig. 1).

If we put the magnetic field along the $z$-direction and write it as $B_3$,
the equations (\ref{13}) might also be written as
\begin{equation}
\label{14}\frac{d{\bf S}_i}{dt}=\left\{ {\bf S}_i,H\right\} =-\frac{geB_3}{%
2mc}\epsilon _{i3k}S_k
\end{equation}
which can be written in vector notation as
\begin{equation}
\label{15}\frac{d{\bf S}_i}{dt}=\frac{ge}{2mc}\left( {\bf S}\wedge {\bf B}%
\right) _i.
\end{equation}
{}From this last equation we might write the dipole moment of the electron as
\begin{equation}
\label{16}{\bf m}_s=\frac{ge}{2mc}{\bf S}.
\end{equation}
This justifies the expression (\ref{3}) if we substitute $S$ for $L_z$ and
the Land\'e $g$-factor for $2$. We then expect from a purelly classical
argument that the structure of the half-integral spin particles be somewhat
like a ring.

We now introduce the functions
\begin{equation}
\label{17}S_{+}=S_1+iS_2\mbox{ ; }S_{-}=S_1-iS_2,
\end{equation}
that are given in terms of the coordinates and momenta as
\begin{equation}
\label{18}S_{+}=\frac 14\left[ \sqrt{\frac \alpha \beta }\left( x+iy\right)
^2+\sqrt{\frac \beta \alpha }\left( p_x+ip_y\right) ^2\right] ,
\end{equation}
and
\begin{equation}
\label{19}S_{-}=\frac 14\left[ \sqrt{\frac \alpha \beta }\left( x-iy\right)
^2+\sqrt{\frac \beta \alpha }\left( p_x-ip_y\right) ^2\right] .
\end{equation}
These functions satisfy the following commutation rules
\begin{equation}
\label{19.a}\left\{ S_{+},S_{-}\right\} =-2S_3\mbox{ ; }\left\{
S_{+},S_{+}\right\} =\left\{ S_{-},S_{-}\right\} =0,
\end{equation}
while for the anti-commutator, defined classically as
\begin{equation}
\label{19.b}\{f,g\}_A=\sum_{k=1}^3\left( \frac{\partial f}{\partial q_k}
\frac{\partial g}{\partial p_k}+\frac{\partial f}{\partial p_k}\frac{%
\partial g}{\partial q_k}\right) ,
\end{equation}
we get the anti-commutation rules

\begin{equation}
\label{20}\left\{ S_{+},S_{-}\right\} _A=(xp_x+yp_y)\mbox{ ; }\left\{
S_{+},S_{+}\right\} _A=\left\{ S_{-},S_{-}\right\} _A=0.
\end{equation}
We can easily identify the element $xp_x+yp_y$ as the unit associated with
the product defined by the anti-commutation relation if we note that, for
this element
\begin{equation}
\label{21.0}\left\{ S_{+},xp_x+yp_y\right\} _A=2S_{+}\mbox{ ; }\left\{
S_{-},xp_x+yp_y\right\} _A=2S_{-}
\end{equation}
and then write the first expression in (\ref{20}) formally as
\begin{equation}
\label{21.1}\left\{ S_{+},S_{-}\right\} _A={\bf 1}.
\end{equation}

We might also obtain another important element of description of our
problem. Defining the function $N$ as the product
\begin{equation}
\label{21a}N=S_{+}S_{-}=S_1^2+S_2^2
\end{equation}
it is easy to see that
\begin{equation}
\label{21b}N=S^2-S_3^2=-\frac 14L_z^2+\frac 14\left[ \left( \frac 12\right)
\sqrt{\frac \alpha \beta }\left( x^2+y^2\right) +\left( \frac 12\right)
\sqrt{\frac \beta \alpha }.\left( p_x^2+p_y^2\right) \right] ^2
\end{equation}
which is nothing but the Casimir operator in this coordinate-momentum
representation of the Lie group SU(3) generated by the three functions $S_i$
(with $z=p_z=0$) and other five ones that will not concern us here.

The physical meaning of the relations (\ref{19.a},\ref{20}) together with
the functions (\ref{18}), (\ref{19}) and (\ref{21a}) will be clarified in
the next section.

Until now we were using the passive approach. In this approach we view the
particle as a body with some intrinsic features moving on a fixed background
or coordinate system. These features, like its rotation around the $z$-axis,
are described by the {\it functions} $S_i$ (e.g. for the rotation around the
$z$-axis we have $S_3$). We now pass to the active point of view.

\section{The Active View: Operators}

In the active approach, the particle is neither rotating nor being distorted
by some applied magnetic field. In this case, the particle is maintained
fixed in space and is viewed as generating symmetry operations upon the
space itself (e.g. the electron generates the rotation along the $z$-axis
and this operation is now represented by the operator $\hat L_z$).

Since we know that the functions $S_i$ are the generators, in the
coordinate-momentum representation, of the SU(2) symmetry group, the
operators related to them are automatically obtained (for the half-spin
case) as the Pauli operators, written in matrix form as%
\footnote{The reader may, at this point, say that we have
forced this result through our choice of the $S_i$'s functional appearance
and that this implies a great degree of arbitrariness. The reader is correct
but we might also complain that the same is done in the usual quantum
mechanical treatment when one chooses the $\hat{L}_z$ operator to be
$\sigma_z$, based in the results obtained in the Stern-Gerlach experiment and,
after
that, derive the other two Pauli operators $\sigma_x$ and $\sigma_y$.
In both cases, of course, one is driven by the known behavior of the electron
under the influence of a uniform magnetic field. The classical path is
different from the quantum one just in the sense that it claims for a picture
of the electron, something the quantum procedure does not.}
$$
\widehat{\sigma }_1=\left(
\begin{array}{cc}
0 & 1 \\
1 & 0
\end{array}
\right) \mbox{ ; }\widehat{\sigma }_2=\left(
\begin{array}{cc}
0 & -i \\
i & 0
\end{array}
\right) \mbox{ ; }\widehat{\sigma }_3=\left(
\begin{array}{cc}
1 & 0 \\
0 & -1
\end{array}
\right)
$$
so that our $S$-functions shall be proportional to them and might be
written, in this matrix representation, as
\begin{equation}
\label{22}\hat S_3=\frac h2\widehat{\sigma }_3\mbox{ ; }\hat S_2=-i\frac h2
\widehat{\sigma }_2\mbox{ ; }\hat S_1=\frac h2\widehat{\sigma }_1,
\end{equation}
where $h$ is an yet undetermined constant that will be obtained from the
experiments (e.g. the Stern-Gerlach experiment) and is indeed known to be
Plank's constant $\hbar $---which we will hereafter write instead of $h$.

This definition of the $S_i$ functions also assures that we have obeyed the
usual commutation relation
\begin{equation}
\label{22.a}\left[ \hat S_i,\hat S_j\right] =i\hbar \epsilon _{ijk}\hat S_k.
\end{equation}

The hamiltonian related to the interaction of the particle with an external
magnetic field becomes
\begin{equation}
\label{23}H=-\frac{ge\hbar }{2mc}B_3\frac{\widehat{\sigma }_3}2.
\end{equation}
Using expression (\ref{17}) we get

\begin{equation}
\label{24}\hat S_{+}=\frac \hbar 2\left(
\begin{array}{c}
0 \\
1
\end{array}
\begin{array}{c}
0 \\
0
\end{array}
\right)
\end{equation}
and
\begin{equation}
\label{25}\hat S_{-}=\frac \hbar 2\left(
\begin{array}{c}
0 \\
0
\end{array}
\begin{array}{c}
1 \\
0
\end{array}
\right) ,
\end{equation}
with the commutation relations---in terms of the matrix commutator---given by

\begin{equation}
\label{26}\left[ \hat S_{+},\hat S_{-}\right] =\hbar \hat S_3\mbox{ ; }%
\left[ \hat S_{+},\hat S_{+}\right] =\left[ \hat S_{-},\hat S_{-}\right] =0,
\end{equation}
while the anti-commutation relations are

\begin{equation}
\label{27}\left[ \hat S_{+},\hat S_{-}\right] _A=\hbar {\bf 1}\mbox{ ; }%
\left[ \hat S_{+},\hat S_{+}\right] _A=\left[ \hat S_{-},\hat S_{-}\right]
_A=0.
\end{equation}

Since the operators $\hat S_{+}$ and $\hat S_{-}$ act as rotations upon the
two-dimensional space defined by the operators $\hat S_1$ and $\hat S_2$ we
might define, over this space, the basis
\begin{equation}
\label{28}|0\rangle =\left(
\begin{array}{c}
1 \\
0
\end{array}
\right) \mbox{ ; }|1\rangle =\left(
\begin{array}{c}
0 \\
1
\end{array}
\right) ,
\end{equation}
chosen as to make the representation of $\hat S_3$ and $\hat S^2$ diagonal.
In this case it is easy to see that the operators $\hat S_{+}$ and $\hat
S_{-}$ are the normal modes associated with the distortions of the particle
in the $xyp_xp_y$-hyperplane referred to after expression (\ref{15}).

It is then easy to see that we have
\begin{equation}
\label{29} \hat{S}_{+}|0\rangle=|1\rangle\mbox{ ; } \hat{S}_{+}|1\rangle=0%
\mbox{ ; } \hat{S}_{-}|0\rangle=0\mbox{ ; } \hat{S}_{-}|1\rangle=|0\rangle.
\end{equation}

We might now modify our terminology of the operator $\hat S_{\pm }$ from
symmetry generators to operators related with particles and occupation
numbers. In this case, $|0\rangle $ is the vacuum state and $|1\rangle $ the
first occupation state and $\hat S_{+}$ behaves as a particle creation
operator while $\hat S_{-}$ behaves as a particle annihilation operator.
This terminology, however, might be misleading. The operators $\hat S_{+}$
and $\hat S_{-\mbox{ }}$are only entities allowing one to get one
independent mode of distortion in the $xyp_xp_y$-plane from the other. There
are no particles being created nor annihilated.

By using the particle terminology and looking at expression (\ref{29}) we
might say that it is not possible to find any state with more than one
particle. It is easy to define a number operator given by
$$
\hat N=\hat S_{+}\hat S_{-}
$$
with matrix representation written as
\begin{equation}
\label{30}\hat N=\left(
\begin{array}{c}
0 \\
0
\end{array}
\begin{array}{c}
0 \\
1
\end{array}
\right) ,
\end{equation}
for which

\begin{equation}
\label{31}\hat N|0\rangle =0\mbox{ ; }\hat N|1\rangle =+1|1\rangle ,
\end{equation}
justifying the name just given to it.

Returning to the expression (\ref{21a}), now written in operator format,
\begin{equation}
\label{32}\stackrel{\symbol{94} }{N}=\stackrel{\symbol{94} }{S}_{+}\stackrel{%
\symbol{94} }{S}_{-}=\stackrel{\symbol{94} }{S}_1^2+\stackrel{\symbol{94} }{S%
}_2^2=\stackrel{\symbol{94} }{S}^2-\stackrel{\symbol{94} }{S}_3^2
\end{equation}
we might also interpret this operator $\stackrel{\symbol{94} }{N}$ as
representing the net `angular momentum' related with the distortions. Then,
the result
\begin{equation}
\label{33}\hat N|0\rangle =0
\end{equation}
is related with the fact that no distortion is present, while the result
\begin{equation}
\label{34}\hat N|1\rangle =+1|1\rangle
\end{equation}
implies that the particle is being distorted. The condition that $0$ and $1$
are the only possible eigenvalues of the operator $N$ means that the
particle is capable of being in only two states---one with and another
without distortion in the $xyp_xp_y$-hyperplane.

We can also explain why the particle symmetry group is SU(2). The functions $%
S_1$ and $S_2$ or their related operators are not true angular momenta but
rather components of a tensor. In this case it is known that when we perform
rotations upon the coordinates by an angle $\Omega $ around their symmetry
axis, using the rotation group of which $\Omega $ is the sole parameter,
these tensors are transformed by an angle $\Omega /2$ and so, are related
with the covering group SU(2) of O(3).

\section{Pauli's Exclusion Theorem}

Since the beginning of quantum mechanics the behavior of exclusion presented
by electrons (and all half-integral spin particles) is introduced into
non-relativistic quantum mechanics by means of a Principle, an axiom, called
the Exclusion Principle due to Pauli. It is amply believed that the correct
place to fit this problem is relativistic quantum mechanics where Dirac's
matrices arrive naturally and obey anticommutation relations. From the point
of view of this paper this is not the case.

The spin, it was shown is not a relativistic effect although it is more
correctly addressed within the realm of this theory---as all other
phenomena. In relativistic quantum mechanics the spin shall have other
degrees of freedom, for questions of invariance, related with the coupling
of these degrees of freedom with the electric field. This is represented by
the Lorentz invariant potential\cite{eu3-5}
\begin{equation}
\label{35}L=\Sigma \cdot {\bf H}+\Xi \cdot {\bf E}
\end{equation}
where $\Sigma $ and $\Xi $ are the Dirac's matrices and ${\bf H}$ and ${\bf E%
}$ are the magnetic and electric fields respectively.

In Dirac's theory, the matrix representation of the spin is related with
4-spinors---in contrast with Pauli's 2-spinors---where all these degrees of
freedom are made explicit. The use of spinors, however, need not be
considered fundamental for the theory. It reflects just a separation of the
problem into two different approaches---one in which the spatial coordinates
are treated analytically, by means of the Schr\"odinger equation, and the
other in which the intrinsic problem is treated by matrix algebra, following
the Heisemberg original approach. Such an approach could be also used in the
non-relativistic hydrogen atom, for example, treating the angular operators $%
L_3$ and $L^2$ as matrices and so obtaining, for each quantum number $\ell $
of operator $L^2$, a $(2\ell +1)$-spinor.

The use of matrix representation for the spin is made because it is tacitly
assumed among the scientific community that the spin cannot be represented
by ordinary analytical functions. This follows from another general faith
that the spin has no classical analogous and so, is not suitable to
quantization. This argument, however, cannot be correct if we accept that
there is a total correspondence between the Schr\"odinger's and Heisemberg's
calculus. In the continuation paper (VII), we will present an analytical
equation accounting for the particles half-integral spin.

For the present section we are interested in showing that the results
presented until now are the only ones necessary to introduce the notion of
exclusion. This means that the behavior of exclusion will be consequence of
the particle geometry which is also responsible for the existence of spin.

To begin with, let us consider two independent half-spin particles described
by the two sets of functions
\begin{equation}
\label{36}\left\{
\begin{array}{l}
S_1=\frac 12\left( x_1y_1+p_{x1}p_{y1}\right)  \\
S_2=\frac 14\left( x_1^2-y_1^2+p_{x1}^2-p_{y1}^2\right)  \\
S_3=\frac 12\left( x_1p_{y1}+y_1p_{x1}\right)
\end{array}
\right. \ \mbox{and\ }\left\{
\begin{array}{l}
R_1=\frac 12\left( x_2y_2+p_{x2}p_{y2}\right)  \\
R_2=\frac 14\left( x_2^2-y_2^2+p_{x2}^2-p_{y2}^2\right)  \\
R_3=\frac 12\left( x_2p_{y2}+y_2p_{x2}\right)
\end{array}
\right.
\end{equation}
with the commutation rules
\begin{equation}
\label{37}\left\{ S_i,S_j\right\} =\epsilon _{ijk}S_k\ \mbox{and}\ \left\{
R_i,R_j\right\} =\epsilon _{ijk}R_k
\end{equation}
and
\begin{equation}
\label{38}\left\{ S_i,R_j\right\} =0\mbox{ for all }i,j.
\end{equation}

It is easily to work in the active view with the matrix representation. In
this case we are interested in the tensorial space $S\otimes {\bf 1}+{\bf 1}%
\otimes R$ with the probability amplitudes defined by $\left| s\right\rangle
_S\otimes \left| r\right\rangle _R$. It has to be stressed that we are
looking for a set of vectors making  simultaneously diagonal the operators $%
S_3+R_3$ and $N_{total}=S_{+}S_{-}+R_{+}R_{-}$.

It can be verified that we have
\begin{equation}
\label{39}S_3+R_3=\left(
\begin{array}{cccc}
2 & 0 & 0 & 0 \\
0 & 0 & 0 & 0 \\
0 & 0 & 0 & 0 \\
0 & 0 & 0 & -2
\end{array}
\right) \mbox{ and }N_{tot}=\left(
\begin{array}{cccc}
0 & 0 & 0 & 0 \\
0 & 1 & 0 & 0 \\
0 & 0 & 1 & 0 \\
0 & 0 & 0 & 2
\end{array}
\right)
\end{equation}
and we shall choose some combinations of the vectors $\left| s\right\rangle
_S\otimes \left| r\right\rangle _R$ with $s=0,1$ and $r=0,1$ obeying
\begin{equation}
\label{40}\left\{
\begin{array}{l}
S_3\left| 0\right\rangle _S\left| 0\right\rangle _R=\frac 12\left|
0\right\rangle _S\left| 0\right\rangle _R \\
S_3\left| 0\right\rangle _S\left| 1\right\rangle _R=\frac 12\left|
0\right\rangle _S\left| 1\right\rangle _R \\
S_3\left| 1\right\rangle _S\left| 0\right\rangle _R=-\frac 12\left|
1\right\rangle _S\left| 0\right\rangle _R \\
S_3\left| 1\right\rangle _S\left| 1\right\rangle _R=-\frac 12\left|
1\right\rangle _S\left| 1\right\rangle _R
\end{array}
\right. \mbox{ and }\left\{
\begin{array}{l}
R_3\left| 0\right\rangle _S\left| 0\right\rangle _R=\frac 12\left|
0\right\rangle _S\left| 0\right\rangle _R \\
R_3\left| 0\right\rangle _S\left| 1\right\rangle _R=-\frac 12\left|
0\right\rangle _S\left| 1\right\rangle _R \\
R_3\left| 1\right\rangle _S\left| 0\right\rangle _R=\frac 12\left|
1\right\rangle _S\left| 0\right\rangle _R \\
R_3\left| 1\right\rangle _S\left| 1\right\rangle _R=-\frac 12\left|
1\right\rangle _S\left| 1\right\rangle _R
\end{array}
\right.
\end{equation}
that are still eigenvectors of the operators defined in expression (\ref{39}%
).

It is easy to verify that the only two possible combinations that are still
eigenvectors of the first operator in (\ref{39}) are the two symmetrized
vectors
\begin{equation}
\label{41}\left| \psi _1\right\rangle =\frac 1{\sqrt{2}}\left( \left|
0\right\rangle _S\left| 1\right\rangle _R-\left| 1\right\rangle _S\left|
0\right\rangle _R\right) \mbox{ and }\left| \psi _2\right\rangle =\frac 1{
\sqrt{2}}\left( \left| 0\right\rangle _S\left| 1\right\rangle _R+\left|
1\right\rangle _S\left| 0\right\rangle _R\right)
\end{equation}
and have both zero as their eigenvalue as related to operator $S_3$. The
negative sign is then chosen using the anticommutation rule related with
operators $S_{+},S_{-},R_{+}$ and $R_{-}$ \cite{Schiff}. The only solution
for the problem becomes
\begin{equation}
\label{42}\left| \psi _1\right\rangle =\frac 1{\sqrt{2}}\left( \left|
0\right\rangle _S\left| 1\right\rangle _R-\left| 1\right\rangle _S\left|
0\right\rangle _R\right)
\end{equation}
as expected. In matrix notation this vector can be written as
\begin{equation}
\label{43}\left| \psi _1\right\rangle =\left(
\begin{array}{c}
0 \\
1 \\
-1 \\
0
\end{array}
\right)
\end{equation}
which is a statement of the exclusion behavior.

We then conclude that fixing the geometry of the particle structure, which
induces the anticommutation relations for the operators $S_{+},S_{-},R_{+}$
and $R_{-}$, and the need to make diagonal the operator $S_3+R_3$ are
sufficient to obtain the exclusion behavior of entities with half-integral
spin---we will prove in paper VII of this series that all half spin
particles are included in our calculation.

\section{The Bose-Einstein Condensation}

The picture for the electron made above has other secondary, but not
unimportant, consequences. One of the most striking ones is related with
Einstein-Bose condensation. This phenomenon is related to the disappearance
of exclusion behavior for very low temperatures.

In this paper we will address this problem qualitatively and show that we
expect such phenomenon to appear. In the continuation paper (VII), where the
half-integral spin Schr\"odinger equation will be solved exactly, the
problem will be dealed with by quantitative calculations.

The most important thing to stress is that the internal energy of the
electron is given if we consider the function
\begin{equation}
\label{EB1}S_0=\frac 12\left[ \sqrt{\frac \alpha \beta }\left(
x^2+y^2\right) +\sqrt{\frac \beta \alpha }\left( p_x^2+p_y^2\right) \right]
\end{equation}
and remember that it will give rise, in Schr\"odinger representation, to an
eigenvalue equation
\begin{equation}
\label{EB2}\widehat{S}_0\psi =\hbar \lambda \psi ,
\end{equation}
where Planck's constant was introduced to make $\lambda $ a number without
dimension.

Equation (\ref{EB2}) might be written, after quantized, as
\begin{equation}
\label{EB3}\frac 12\left[ \sqrt{\frac \alpha \beta }\left( \widehat{x}^2+
\widehat{y}^2\right) +\sqrt{\frac \beta \alpha }\left( \widehat{p}_x^2+
\widehat{p}_y^2\right) \right] \psi =\hbar \lambda \psi .
\end{equation}
It might be easily checked that the ration $\sqrt{\alpha /\beta }$ shall
have dimensions of mass/time and we can represent it as
\begin{equation}
\label{EB4}\sqrt{\frac \alpha \beta }=m\omega ,
\end{equation}
where $m$ is the mass of the particle and $\omega $ is some frequency
related to the specific particle under consideration. Equation (\ref{EB3})
becomes
\begin{equation}
\label{EB5}\left[ \frac 1{2m}\left( \widehat{p}_x^2+\widehat{p}_y^2\right)
+\frac 12m\omega ^2\left( \widehat{x}^2+\widehat{y}^2\right) \right] \psi
=\hbar \omega \lambda \psi ,
\end{equation}
with a related {\em internal} energy given by
\begin{equation}
\label{EB6}E=\lambda \hbar \omega .
\end{equation}

Equation (\ref{EB6}) is our fundamental equation. It is noteworthy that we
do not expect the quantum number $\lambda $ to have zero as one of its
possible values. This is because we admitted from the very beginning that
the particle is spinning around some fixed direction and so, must have some
internal energy---this is a striking difference from fermions to bosons. In
this case we expect
\begin{equation}
\label{EB7}\lambda \geq N>0,
\end{equation}
for some constant $N$.

Let us suppose now that we have a confined fermion gas and we begin to drop
its temperature. The fermions energy is the sum of their translation energy
plus their internal energy. The energy is being extracted from the fermions
translational kinetic energy. This process continues until the temperature
reaches a value such that one fermion looses not only its translation energy
but also its internal energy. We then say that the fermion `freezes'.

When the fermion freezes, its internal energy is under the minimum value $%
N\hbar \omega $ and the equation (\ref{EB2}) will not have any solution. The
operator $S^2$ defined above will not define an eigenstate either. We might
then say that the group generated by the fermion will not be SU(2) anymore.
It will no longer be a fermion and condensation will take place---it is like
if the possibility for the fermion to rotate and deform assures the
exclusion; when the temperature drops to an exceeding low value, the
internal state related with these degrees of freedom of the internal
movements is no longer allowed and the fermion freezes---it is important to
stress that this is not a collective behavior.

All the considerations above will be approached quantitatively in the
continuation paper (VII).

\section{Conclusions}

In this paper we have shown how to make a model of the half-integral spin
particles in the realm of a classical theory. It was also shown how the
passage from the passive to the active views might introduce all the
`quantum mechanical' results.

These results then show, for those who believe in some abyss between
classical and quantum physics, that the half-integral spin particles are not
specific of quantum mechanics.

As one consequence of the present development, it was possible to present an
explanation of the intriguing phenomenon of Bose-Einstein condensation for
fermions. We are now in position to `quantize' the functions $S_3$ and $S^2$
to obtain a Schr\"odinger-like representation---eigenfunctions---for the
half-integral spin particles.

In the continuation paper VII we proceed to make the quantum calculations
and to treat the Bose-Einstein condensation problem in a quantitative way.

\section{Acknowledgements}

The author wishes to thanks the Conselho Nacional de Desenvolvimento
Cient\'ifico e Tecnol\'ogico (CNPq) for sponsoring this research.

\end{document}